# Role of Elastic Phonon Couplings in Dictating the Thermal Transport across Atomically Sharp SiC/Si Interfaces

Qinqin He[1#], Yixin Xu[1#], Haidong Wang[2], Zhigang Li and Yanguang Zhou[1]*

[1]*Department of Mechanical and Aerospace Engineering, The Hong Kong University of Science and Technology, Clear Water Bay, Kowloon, Hong Kong SAR*

[2]*Key Laboratory of Thermal Science and Power Engineering of Ministry of Education, Tsinghua University, Beijing, 100084, China*

## Abstract

Wide-bandgap (WBG) semiconductors have promising applications in power electronics due to their high voltages, radio frequencies, and tolerant temperatures. Among all the WBG semiconductors, SiC has attracted attention because of its high mobility, high thermal stability, and high thermal conductivity. However, the interfaces between SiC and the corresponding substrate largely affect the performance of SiC-based electronics. It is therefore necessary to understand and design the interfacial thermal transport across the SiC/substrate interfaces, which is critical for the thermal management design of these SiC-based power electronics. This work systematically investigates heat transfer across the 3C-SiC/Si, 4H-SiC/Si, and 6H-SiC/Si interfaces using non-equilibrium molecular dynamics simulations and diffuse mismatch model. We find that the room temperature ITC for 3C-SiC/Si, 4H-SiC/Si, and 6H-SiC/Si interfaces is 932 MW/m$^2$K, 759 MW/m$^2$K, and 697 MW/m$^2$K, respectively. We also show the contribution of the ITC resulting from elastic scatterings at room temperature is 80% for 3C-SiC/Si interfaces, 85% for 4H-SiC/Si interfaces, and 82% for 6H-SiC/Si interfaces, respectively. We further find the ITC contributed by the elastic scattering decreases with the temperature but remains at a high ratio of 67%~78% even at an ultrahigh temperature of 1000 K. The reason for such a high elastic ITC is the large overlap between the vibrational density of states of Si and SiC at low frequencies (< ~ 18 THz), which is also demonstrated by the diffuse mismatch mode. It is interesting to find that the inelastic ITC resulting from the phonons with frequencies higher than the cutoff frequency of Si (i.e., ~18 THz) can be negligible. That may be because of the wide frequency gap between Si and SiC, which makes the inelastic scattering among these phonons challenging to meet the energy and momentum conservation rules. This work reveals the intrinsic thermal transport mechanism between WBG SiC and Si, which benefits the design of SiC-based power electronics.

---

[#]These authors contribute equally to this work. *Author to whom all correspondence should be addressed. Email: maeygzhou@ust.hk (Y. Zhou)

# 1. Introduction

Wide-bandgap (WBG) semiconductor SiC has promising applications in power electronics due to its high breakdown voltage and radio frequency[1]. However, the self-heating caused by the current flow during the operation of these power electronics increases the corresponding working temperature, reducing the carrier's mobility and the threshold voltage[2]. The heat dissipation capacity of the components (e.g., SiC) in these power electronics, which is strongly related to their thermal properties, is therefore critical for the corresponding performance of these electronics. For instance, the size of 3C-SiC-based metal-oxide-semiconductor field-effect transistors (MOSFETs) is 20 times smaller than that of Si-based MOSFETs when the power density is similar[3]. That is mainly because the thermal conductivity of 3C-SiC (i.e., ~500 W/mK at room temperature[4]) is much higher than that of Si (i.e., ~150 W/mK at room temperature[5]).

The SiC in MOSFETs is usually attached to the substrates (e.g., Si substrate), and the heat dissipation capacity of the corresponding MOSFETs is, therefore, affected by the interfaces between the semiconductor and the substrate. For example, a previous study showed that the maximum temperature of $\beta$-Ga$_2$O$_3$/substrate heterostructures decreased from 1860 K to 1153 K when the interfacial thermal conductance (ITC) of $\beta$-Ga$_2$O$_3$/substrate heterostructures increased from 10 MW/m$^2$K to 110 MW/m$^2$K [6]. While the SiC has high thermal conductivities, as mentioned above[3,4], the heat capacity of the SiC-based electronics may not be as expected since the thermal transport across SiC/substrate interfaces may be poor. It is known that the ITC of SiC-substrate interfaces is determined by the type of substrates[4,7,8], interfacial bond strength[9], and interfacial morphology[10]. For instance, the ITC at room temperature between 4H-SiC and graphene is 19±1.65 × 10$^3$ W/m$^2$K[8], while the room temperature ITC between 3C-SiC and Si can be as high as 622±123 MW/m$^2$K[4]. For another example, Xu. *et al*. [10] found that the room temperature ITC of 4H-SiC/Si interfaces can be changed from ~300 MW/m$^2$K to ~1000 MW/m$^2$K when nanopatterns are introduced in the interfacial region. To design the SiC/substrate heterostructures with better heat dissipation performance, it is therefore critical to understand the thermal transport across the interfaces between SiC and the substrate.

In the past decades, many researchers have attempted to quantify the thermal transport across interfaces experimentally[11–13]and numerically[14–17]. For instance, Hohensee. *et al.*[18] showed that the inelastic thermal conductance across metal/diamond interfaces is dominated by partial transmission processes. Cheng *et al.*[19] experimentally found that the interfacial vibrational modes in Si/Ge interfaces could contribute ~5% to the total ITC. Li *et al.*[20] measured the ITC of Al/Si interfaces and found that the inelastic couplings could contribute largely (i.e., 30% - 48% quantified by Xu *et al.*[21]) to the total ITC in the temperature range of 300~700 K. In the framework of equilibrium molecular dynamics (EMD) simulations, Gordiz and Henry[22] showed that the higher overlap of vibrational density of state (VDOS) of two contact leads did not guarantee higher ITC. They further found that the anharmonicity which activates these high-order scattering channels at interfaces will largely benefit the interfacial thermal transport[22], which is also demonstrated by Xu *et al*.[21], Feng and Ruan[23], Zhou and Hu[15], Kimmo and Volz[24] and Wu and Luo[25]. While previous measurements and simulations have largely advanced our understanding of the thermal transport across the interfaces, the assumptions used to deduce the conclusions in experiments (e.g., using diffuse mismatch model [26] to consider the elastic ITC) and inherent Boltzmann distribution of phonons in MD simulations, strongly affect the accuracy and transferability of the results.

In this paper, we quantitatively and systematically investigate the thermal transport spectrum across the 3C-SiC/Si, 4H-SiC/Si, and 6H-SiC/Si interfaces in the framework of non-equilibrium molecular dynamics (NEMD) simulations while considering quantum effects. The room temperature ITC for 3C-SiC/Si, 4H-SiC/Si, and 6H-SiC/Si interfaces is found to be 932.5 MW/m2K, 696.8 MW/m2K and 758.7 MW/m2K, respectively. Furthermore, the ITC contributed by the elastic couplings at interfaces decreases with the temperature but remains at an ultrahigh ratio of 67% - 80% when the temperature ranges from 900 K to 1200 K. The high overlap of VDOS between Si and SiC with frequencies lower than 18 THz is found to be responsible for this ultrahigh elastic ITC, which is further demonstrated by the diffuse mismatch model. It is also interesting to find that the inelastic ITC contributed by the phonons with frequencies higher than the cutoff frequency of Si can be negligible. That may be because of the wide frequency gap between Si and SiC, which makes the inelastic scatterings among these phonons challenging to meet the scattering rule.

## 2. Methods and simulation details

### 2.1. Details of NEMD simulations

The schematic of the models used in our simulations is shown in **Figure 1a**, in which SiC and Si are located on two sides. Periodical boundary conditions are applied along *x* and *y* directions. The crystal orientations of Si, 3C-SiC, 4H-SiC, and 6H-SiC along the *z* direction are [1 1 1], [1 1 1], [0 0 1] and [0 0 1], respectively. NEMD simulations based on Fourier's law are then used to calculate the thermal transport properties through the open-source application Large-scale Atomic/Molecular Massively Parallel Simulator[27]. The length of 3C-SiC/Si, 4H-SiC/Si, and 6H-SiC/Si interfaces is chosen as 135 nm, 92 nm, and 101 nm, respectively, to eliminate the size effect in our NEMD simulations (**Figure 2**). All the cross-section areas for 3C-SiC/Si, 4H-SiC/Si, and 6H-SiC/Si interfaces are $6.2 \times 5.3$ nm$^2$. The interatomic interactions are described by the Tersoff potential[28]. All the models are first relaxed in the *NPT* ensemble (constant particle number, pressure, and temperature) for 0.125 ns with a timestep of 0.25 fs to eliminate the residual stress in the systems and then switched into *NVT* ensemble (constant particle number, volume, and temperature) for another 0.125 ns. The temperature difference between the hot and cold reservoir is set as 0.5$T$, in which $T$ is the system's temperature. Therefore, the hot and cold reservoir temperatures are 1.25$T$ and 0.75$T$, respectively. The steady temperature distribution is built in the systems after 2.5 ns (**Figure 1b**), and the following 0.5 ns running is used to calculate the ITC based on Fourier's law: $ITC = -Q/(A \cdot \Delta T)$, where $Q$ is the heat current, $A$ is the cross-section area and $\Delta T$ is the temperature drop at interfaces. The heat current $Q$ is calculated by $Q = \partial E/\partial t$ (**Figure 1c**), in which $E$ is the accumulative energy of the heat reservoirs. The temperature drop $\Delta T$ at interfaces is determined by linearly fitting the temperature distributions at two contact sides.

## 2.2 Spectral interfacial thermal transport properties

To quantify the interfacial thermal transport, the spectral interfacial heat current has also been calculated by utilizing the frequency domain direct decomposed method (FDDDM) [15,21,24,29]

$$Q(\omega) = 2 \sum_{i \in Left} \sum_{j \in Right} \int_0^{+\infty} \left\langle \left. \frac{\partial U_j}{\partial \vec{r}_{ji}} \right|_\tau \vec{v}_i(0) - \left. \frac{\partial U_i}{\partial \vec{r}_{ij}} \right|_\tau \vec{v}_j(0) \right\rangle e^{i\omega\tau} d\tau, \qquad (1)$$

where the $U$ is the potential energy, $\vec{r}$ is the distance between the atom $i$ and $j$, and $\partial U_j / \partial \vec{r}_{ji}$ can be regarded as the interatomic forces, $\vec{v}_i$ stands for the atomic velocity of atom $i$. All the inputs are updated every 40 steps, i.e., 10 fs. The interfacial phonon transmission function can then be calculated by

$$\Gamma(\omega) = Q(\omega)/k_B \Delta T. \qquad (2)$$

In this case, $k_B$ is the Boltzmann constant. Meanwhile, the high-order scatterings may be ignored at an ultralow temperature (i.e., 50 K), which is only 7% of the Debye temperature of Si. Therefore, $\Gamma_{elastic}(\omega)$ calculated using **Eq. (2)** at 50 K, it can be regarded as the elastic phonon transmission function.

At the same time, it is found that the ITC based on NEMD simulations overestimates compared to the experimental measurements[21] (**Figure 3**). It is known that the Debye temperatures for Si, 3C-SiC, 4H-SiC, and 6H-SiC are 645 K[30], 1200 K, 1300 K, and 1200 K[31], respectively. Therefore, all the phonons in MD simulations follow the classical limit of Boltzmann distribution and thus are inherently fully occupied [32,33], which is against the real Bose-Einstein distribution of phonons. The Boltzmann distribution can only approximate the Bose-Einstein distribution when the temperature is much higher than the corresponding Debye temperature. Therefore, the ITC calculated using NEMD simulations should consider the quantum correction. Here, we also correct the ITC using the real Bose-Einstein distribution through[21]

$$ITC_{BE}(T) = \frac{1}{2\pi A} \int_0^\infty \hbar\omega \frac{\partial f_{BE}(\omega,T)}{\partial T} \Gamma(\omega,T) d\omega, \qquad (3)$$

where $\hbar$ is the reduced Planck constant, $f_{BE}(\omega,T)$ is the Bose-Einstein distribution, and $\Gamma(\omega,T)$ is the temperature-dependent phonon transmission function calculated from **Eq. (2)**. In addition, the elastic ITC can also be estimated by

$$ITC_{Elastic} \xrightarrow{T = 50\ K} \frac{1}{2\pi A} \int_0^\infty \hbar\omega \frac{\partial f_{BE}(\omega,T)}{\partial T} \Gamma(\omega,T) d\omega. \qquad (4)$$

Then, the inelastic ITC can be obtained using the difference between **Eq. (3)** and **Eq. (4)**.

**2.3 Diffuse mismatch model**

To investigate the influence of the overlap of VDOS on the interfacial thermal transport, we also calculate the ITC between Si and SiC using DMM. Unlike the acoustic mismatch model (AMM), which assumes the phonon propagations as waves and would not scatter at interfaces[34], DMM assumes all phonons diffusively scatter at interfaces[35]. The heat flux across the interface from side 1 to side 2 is given by[26]

$$q_z^{1\to 2} = \frac{1}{8\pi^2} \sum_\lambda \int_{k_{x,\lambda,1}} \int_{k_{y,\lambda,1}} \int_{k_{z,\lambda,1}>0} \hbar\omega_{\lambda,1}(k_{\lambda,1}) \xi^{1\to 2} |v_{\lambda,1}(k_{\lambda,1})| f_0 dk_{z,\lambda,1} dk_{y,\lambda,1} dk_{x,\lambda,1}, \qquad (5)$$

where $z$ is the transport direction, $\lambda$ is the polarization, $\xi$ is the phonon transmission coefficient, $v$ is the group velocity at the corresponding side, $f_0$ is the phonon distribution at side 1, which follows the Bose-Einstein distribution, $\vec{k}$ is the wave vector. The transmission coefficient from side 1 to side 2 in **Eq. (5)** can be calculated using DMM through[26]

$$\xi^{1\to 2}(k_1) = \frac{\sum_\lambda \hbar\omega_{\lambda,2} k_{\lambda,2}^2 v_{\lambda,2} f_0 dk_{\lambda,2}}{\sum_\lambda \hbar\omega_{\lambda,2} k_{\lambda,2}^2 v_{\lambda,2} f_0 dk_{\lambda,2} + \sum_\lambda \hbar\omega_{\lambda,1} k_{\lambda,1}^2 v_{\lambda,1} f_0 dk_{\lambda,1}}. \tag{6}$$

The summation of different polarizations means that the DMM inherently allows phonons to scatter to any other polarizations. Combing with Fourier's law, the ITC in DMM can then be obtained by

$$h_{BD}^{1\to 2} = \frac{q_z^{1\to 2}}{T^{1\to 2}} = \frac{1}{8\pi^2} \sum_\lambda \int_{k_{\lambda,1}} \hbar\omega_{\lambda,1}(k_{\lambda,1}) k_{\lambda,1}^2 \xi^{1\to 2} \left| v_{\lambda,1}(k_{\lambda,1}) \right| \frac{\partial f_0}{\partial T} dk_{\lambda,1}, \tag{7}$$

where $T^{1\to 2}$ is the temperature drop across the interface.

## 3. Results and discussions

### 3.1 Interfacial thermal conductance

It is known that interfacial thermal transport should consider both the elastic and inelastic scattering channels [10,15,24,36,37]. At low temperatures, the interfacial thermal transport is dominated by the elastic scattering channels, and the temperature dependence of the corresponding ITC is solely dependent on the phonon occupation number. In MD simulations, all vibrations are intrinsically assumed to be fully occupied[33,38]. Therefore, the ITC calculated using MD simulations should weakly depend on temperature when the thermal energy exchange across the interface is mainly through elastic scattering channels (i.e., at low temperatures). At high temperatures, the inelastic scattering channels are found to contribute to the ITC of Si/Al interfaces by 30%-48%[21]. Even at room temperature, around ~25% of the ITC of the Si/Ge interface is contributed by inelastic scattering channels[15]. The classic AMM and DMM, which only consider the elastic scattering channels at interfaces, are therefore inherently inaccurate in depicting the interfacial thermal transport at high temperatures.

Here, our NEMD simulations show that the ITC of Si/3C-SiC interfaces firstly increases with temperature and then converges to 1183.7 MW/m$^2$K at 600 K. It is noted that the ITC calculated NEMD simulations is still around 929.7 MW/m$^2$K even at an extremely low temperature of 50 K. This is because phonons in MD simulations follow the classical limit of Boltzmann distribution, which implies all phonons are fully occupied. However, phonons are bosons and should follow the Bose-Einstein distribution[38]. As a result, the ITC calculated using MD simulations is overestimated at low temperatures of which phonons should be partially occupied. We further calculate the ITC considering the quantum effect through **Eq. (3)**. It is found that the corrected ITC of Si/3C-SiC interfaces is only 55.5 MW/m$^2$K at 50 K, and close to the values calculated using NEMD simulations when the temperature is around 1000 K (**Figure 3a**). For Si/4H-SiC and Si/6H-SiC interfaces, the ITC computed based on the NEMD simulations shows a similar trend to Si/3C-SiC interfaces (**Figure 3a**). It is interesting to find that the converged ITC of Si/4H-SiC and Si/6H-SiC interfaces at 1000 K are similar and have a value of ~1100 MW/m$^2$K. This may be because of the similar VDOS of Si/4H-SiC and Si/6H-SiC interfaces (**Figures 3e-f**). The ITC at 50 K for Si/4H-SiC and Si/6H-SiC interfaces considering quantum effect is 47.7 MW/m$^2$K and 46.4 MW/m$^2$K, respectively. Therefore, the quantum effect in our calculated ITC using NEMD simulations may be ignored only when the temperature is higher than 1000 K. We further compare the ITC of Si/3C-SiC interfaces with the experimental values (**Figure 3b**). Our calculated room-temperature ITC of Si/3C-SiC interface is higher than that of the experimental measurement. This is because there may be defects on the interface in real samples, which can strongly scatter the phonons and decrease the corresponding ITC.

**3.2 Interfacial thermal transport spectrum**

To further investigate the underlying mechanism behind the interfacial thermal transport, we calculate the transmission coefficient function in the framework of NEMD simulations based on **Eq. (2)**. **Figure 4a** shows that the interfacial transmission function of phonons with frequencies smaller than ~8 THz at Si/3C-SiC interfaces increases with temperature, which means more high-order scattering channels among these phonons at high temperatures are activated. It is also noted that the interfacial transmission function of these middle-frequency phonons (i.e., 10~20 THz) is almost independent of temperature. This may be because the high-order scatterings among these phonons are saturated. At high temperatures (e.g., 600 K), these high-frequency (i.e., larger than 20 THz) phonon scattering channels can also be activated. We further calculate the ITC spectrum based on **Eq. (3)**. At low temperatures, only low-frequency phonons are activated and transport thermal energy (**Figure 4d**). More high-frequency phonons are occupied and transport thermal energy when the temperature increases.

For Si/4H-SiC and Si/6H-SiC interfaces (**Figures 4b** and **4c**), it is found that the interfacial transmission function is generally decreasing compared to that of the Si/3C-SiC interface. This is because the difference of the VDOS between Si and 3C-SiC is smaller than that between Si and 4H-SiC or 6H-SiC. The interfacial transmission function of the Si/4H-SiC interface is similar to that of the Si/6H-SiC interface, which results from the fact that the VODS of 4H-SiC is quite similar to that of 6H-SiC (**Figures 3f**). We also find that the interfacial transmission function of low-frequency (i.e., < ~8 THz) phonons increases largely with temperature for both Si/4H-SiC and Si/6H-SiC interfaces (**Figures 4b** and **4c**). Unlike the Si/4H-SiC interface, the interfacial transmission function of these middle-frequency phonons (i.e., 10~20 THz) of Si/6H-SiC interface decreases with temperature (**Figure 4c**), which is due to the phonon softening caused by the temperature as we can observe in the corresponding VDOS (**Figure 3e**).

Recent studies [21,23,36,39,40] showed that the interfacial modes might have a non-negligible contribution to the interfacial thermal transport. Here, we also quantify the contribution of the interfacial phonon modes to the total ITC. We first characterize the interfacial phonon modes based on the modal participation ratios. The 3C-SiC/Si, 4H-SiC/Si, and 6H-SiC/Si supercells consisting of 6104, 1184, and 1776 atoms with a length of 2.5nm, 2 nm, and 3nm at each side are used to calculate the eigenvalues of the phonon modes. The interfacial phonon modes can be distinguished based on the participation parameter contribution (PPC)[17]

$$PPC_{Interface}(v,\vec{k}) = \frac{\sum_{b \in Interface\ region} |\vec{e}_b(v,\vec{k})|}{\sum_{b \in Entire\ system} |\vec{e}_b(v,\vec{k})|}, \tag{8}$$

$$PPC_{SiC\ or\ Si}(v,\vec{k}) = \frac{\sum_{b \in SiC\ or\ Si} |\vec{e}_b(v,\vec{k})|}{\sum_{b \in Entire\ system} |\vec{e}_b(v,\vec{k})|}, \tag{9}$$

where $\vec{e}_b(v,\vec{k})$ is the eigenvector of phonon mode $(v,\vec{k})$ contributed by the atom $b$, which is calculated using Phonopy[41]. $\sum_{b \in Entire\ system} |\vec{e}_b(v,\vec{k})|$ sums the eigenvector magnitudes for eigenmodes contributed by all atoms in the supercell, $\sum_{b \in Interface\ region} |\vec{e}_b(v,\vec{k})|$ sums the eigenvector magnitudes for eigenmodes contributed by the interfacial atoms, and $\sum_{\beta \in SiC\ or\ Si} |\vec{e}_\beta(v,\vec{k})|$ sums the eigenvector magnitudes contributed by atoms in SiC or Si. The PPC is widely used to classify various vibrational modes on the spatial bias [17]. As suggested by Gordiz and Henry[17], these vibrational modes with $PPC_{interface}(v,\vec{k}) > 0.5$ [17] denote the vibration mainly contributed by atoms within the interface region and hence are regarded as the interfacial vibrational modes. In our calculations, 2 or 3-layer atoms at each side nearest to the interface are selected as interfacial region. Our results show that interfacial vibrational modes are mainly located at high-frequency regions (i.e., 18~30 THz) for all three Si/SiC interfaces (**Figure 5d-f**). All these phonons with frequencies of 18~30 THz contribute only 1.8%~6.6% to the total ITC for the Si/SiC interfaces (**Figures 4d-f**). Thus, we can find that the contribution to ITC from these interfacial vibrational modes in all three Si/SiC interfaces should be small and lower than 6.6%.

### 3.3 Elastic and inelastic phonon scattering channels

Before closing, we quantitatively characterize the elastic and inelastic scattering channels at Si/SiC interfaces through **Eqs. (3)** and **(4)**. For the Si/3C-SiC interface, the ITC contributed by these elastic scattering channels increases with temperature as more scattering channels are activated at higher temperatures (**Figure 6a**). When the temperature is above ~ 900 K, all the phonons with frequencies smaller than 18 THz are fully occupied. Our calculated interfacial transmission function shows that these phonons with frequencies lower than 18 THz are the main scattering channels at the Si/SiC interfaces. Therefore, the ITC of the Si/3C-SiC interface resulted from elastic scattering channels converges to 909 W/m$^2$K at ~900 K. We further find that the elastic scattering channels contribute at least 76% to the total interfacial thermal transport. The ITC resulting from inelastic scattering channels are found to increase and then converge around 254 W/m$^2$K at ~900 K. The increase of the ITC contributed by inelastic scattering channels stems from 1, more high-order inelastic scattering channels are activated at higher temperatures, and 2, high-frequency phonons will be fully occupied at high temperatures.

Meanwhile, it is known that the ITC contributed by the elastic scattering channels is determined by two factors. First, the overlap of the VDOS between two contact leads can be asserted by the DMM[26]. Second, the out-of-equilibrium of the interfacial atoms is due to the local discontinuity at the interface[15,36]. The ITC of the Si/3C-SiC interface calculated using DMM agrees well with our quantified elastic value, which implies that the high elastic ITC mainly results from the high overlap of VDOS between Si and 3C-SiC (**Figure 3f**). It should be noted that all the phonons are assumed to be diffusively scattered at the interface in DMM[26], which may not be suitable for the long wavelength phonons as they are spectrally scattered at the interface[42]. As a result, the DMM-based ITC is slightly lower than the elastic ITC calculated based on NEMD simulations.

Because the main interfacial heat carriers are these phonons with frequencies lower than 18 THz (i.e., the corresponding temperature is 850 K) for the Si/4H-SiC and Si/6H-SiC interfaces, the elastic ITC increases and converges to 737.7 W/m$^2$K at ~900 K for Si/4H-SiC interface and 769.4 W/m$^2$K at ~900 K for Si/6H-SiC interface (**Figures 6b** and **6c**), respectively. The elastic phonon scattering channels contribute at least 67% to the ITC of Si/4H-SiC interface and 72% to the ITC of Si/6H-SiC interface (**Figures 6b** and **6c**), respectively. Our results also show that the total ITC of Si/4H-SiC and Si/6H-SiC interfaces becomes almost temperature-independent when the temperature is higher than 900 K. This is because all the phonons with frequencies smaller than 18 THz, which are the main heat carries for the thermal energy exchange across Si/4H-SiC and Si/6H-SiC interfaces, are fully occupied. Meanwhile, the high-order inelastic scatterings at Si/4H-SiC and Si/6H-SiC interfaces are saturated when the temperature is higher than 900 K. We also compare the ITC calculated using DMM with our elastic ITC extracted from our NEMD simulations. It is found that the DDM-based ITC for both Si/4H-SiC and Si/6H-SiC interfaces is larger than our NEMD-based elastic ITC. As discussed above, the DMM can consider the overlap of VDOS between two contact leads and assume all the interfacial atoms have the same neighbour environment, which ignores the local discontinuity at the interface. Therefore, if the interfacial atoms at interfaces are largely out-of-equilibrium caused by the local discontinuity, the DMM will overestimate the ITC, as we have observed in both Si/4H-SiC and Si/6H-SiC interfaces (**Figure**s **6b** and **6c**).

## 4. Conclusion

In summary, we quantitatively study the interfacial thermal transport in 3C-SiC/Si, 4H-SiC/Si, and 6H-SiC/Si in the framework of NEMD simulations and lattice dynamics. Our results reveal that elastic phonon scatterings contribute significantly to the ITC across SiC/Si interfaces, accounting for ~ 82% of the total ITC at room temperature and ~72% even at a high temperature of 1000 K. This ultrahigh contribution to the total ITC resulted from elastic scatterings is stemming from the high overlap of VDOS between Si and SiC, which is further demonstrated by our DMM results. We also find that these phonons with frequencies smaller than 18 THz (i.e., the cutoff frequency of Si) are the main heat carriers, while the phonons with frequencies higher than 18 THz contribute little to interfacial thermal transport. This is caused by the wide frequency gap between Si and SiC, making it difficult for these high-frequency phonons to meet the inelastic scattering rule. Furthermore, the interfacial vibrational modes are found to contribute a little (< 6 % of the total ITC at room temperature) to interfacial thermal transport across Si/SiC interfaces. Our results provide a fundamental understanding of thermal transport across SiC/Si interfaces, which benefits the design of thermal management strategies for SiC-based power electronics.

**Acknowledgments**

Y.Z. thanks the startup fund (REC20EGR14, a/c-R9246) from Hong Kong University of Science and Technology (HKUST) and the ASPIRE Seed Fund (ASPIRE2022#1) from the ASPIRE League. Y.Z. also thanks the fund from the Natural Science Foundation of Guangdong under Grant No. 306206039025.

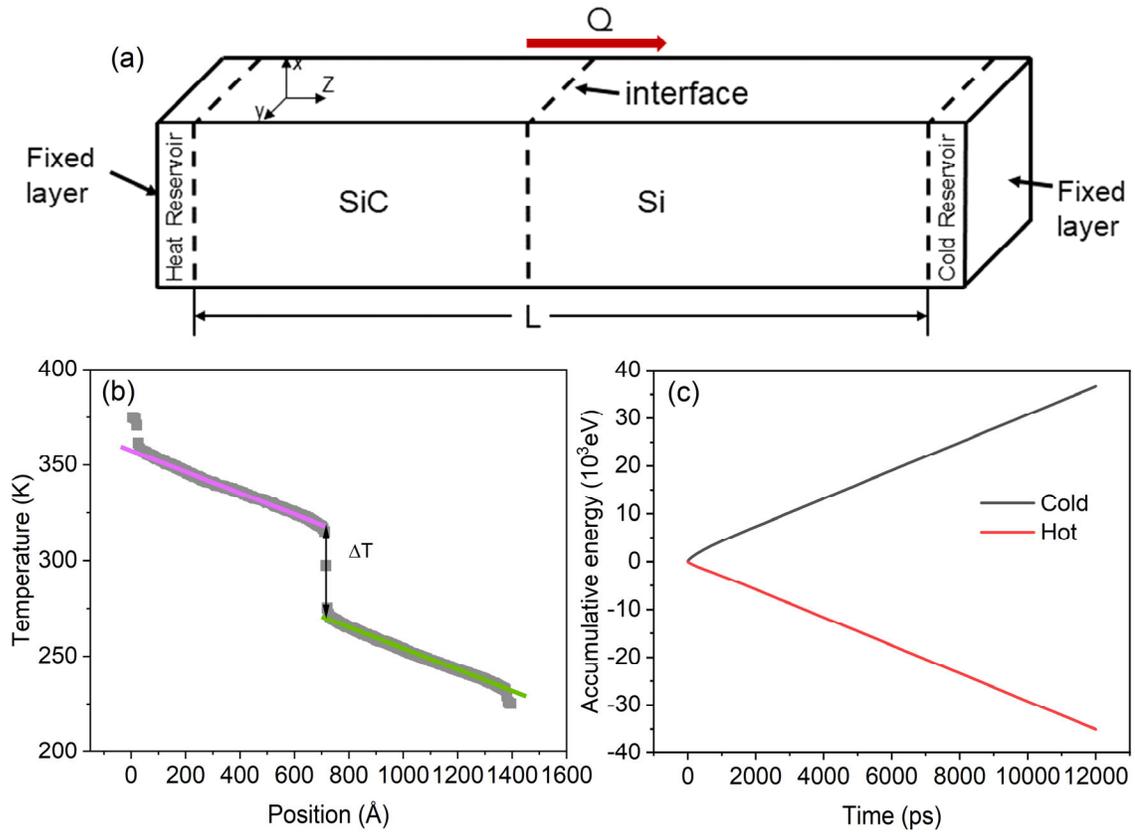

**Figure 1.** (a) The schematic of SiC/Si interfaces. SiC and Si are separated equally, with a heat reservoir added on the SiC side and a cold reservoir added on the Si side. The heat flux is built along *z* direction. (b) The steady temperature distribution along *z* direction at 300K. (c) The accumulative energy of two thermostats during NEMD simulations.

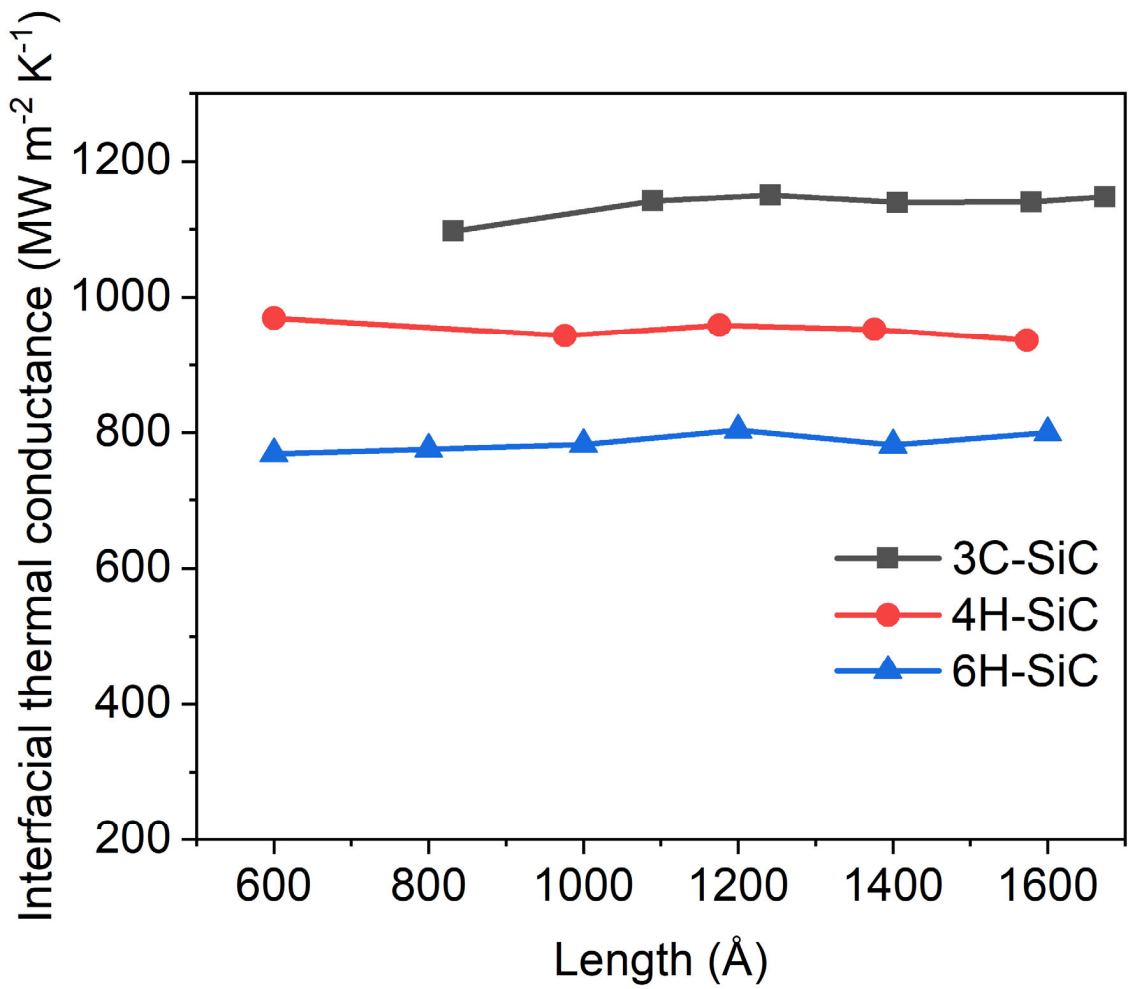

**Figure 2.** The length-dependent interfacial thermal conductance across SiC/Si interfaces in NEMD simulations.

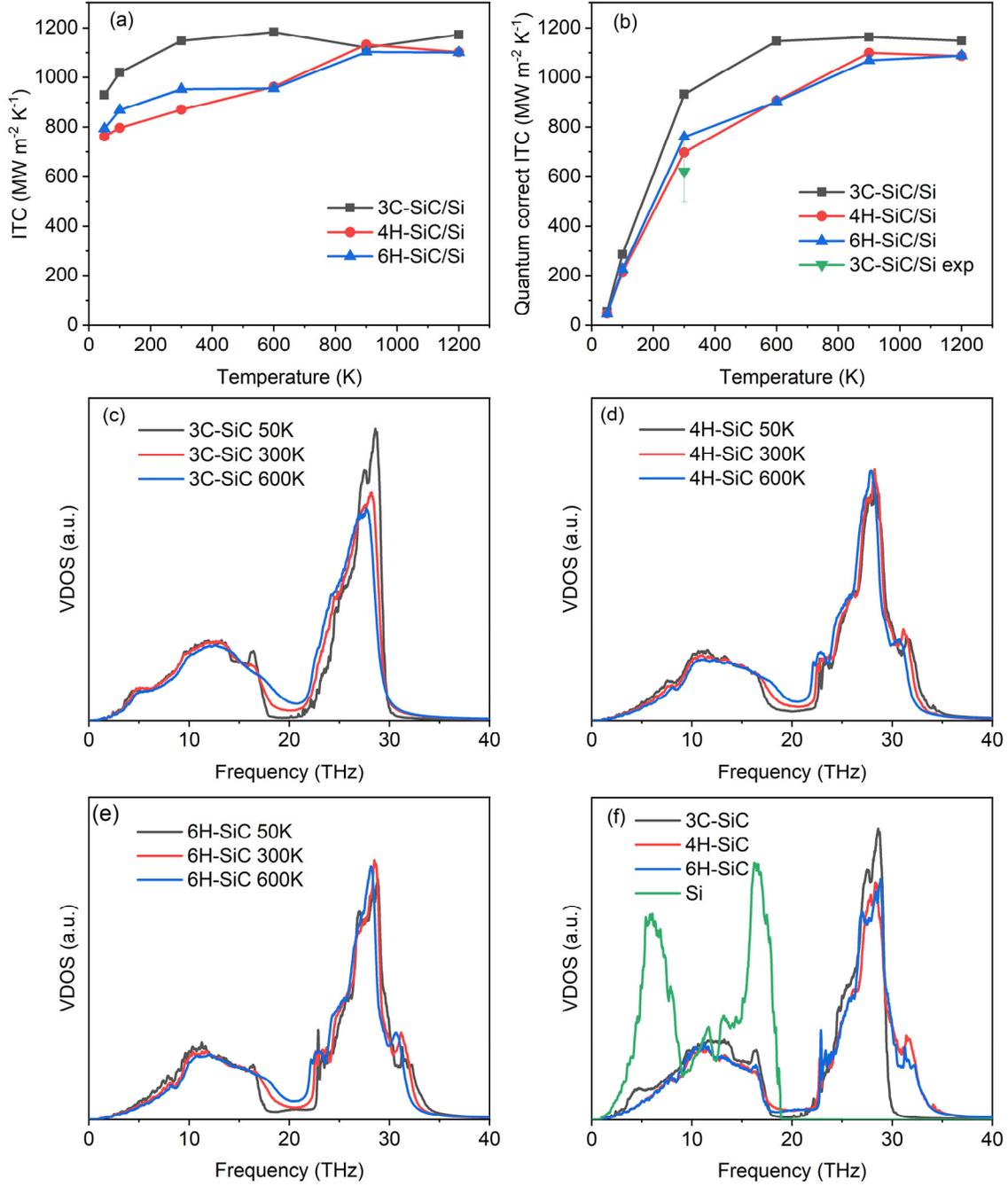

**Figure 3.** (a) The ITC of 3C-SiC/Si, 4H-SiC/Si, and 6H-SiC/Si interfaces calculated using NEMD simulations. (b) The ITC of 3C-SiC/Si, 4H-SiC/Si, and 6H-SiC/Si interfaces after quantum correction and the measured ITC of 3C-SiC/Si[4]. The VDOS at different temperatures of (c) 3C-SiC/Si, (d) 4H-SiC/Si, and (e) 6H-SiC/Si interfaces. (f) The VDOS of Si and 3C-, 4H and 6H-SiC at the interfacial region.

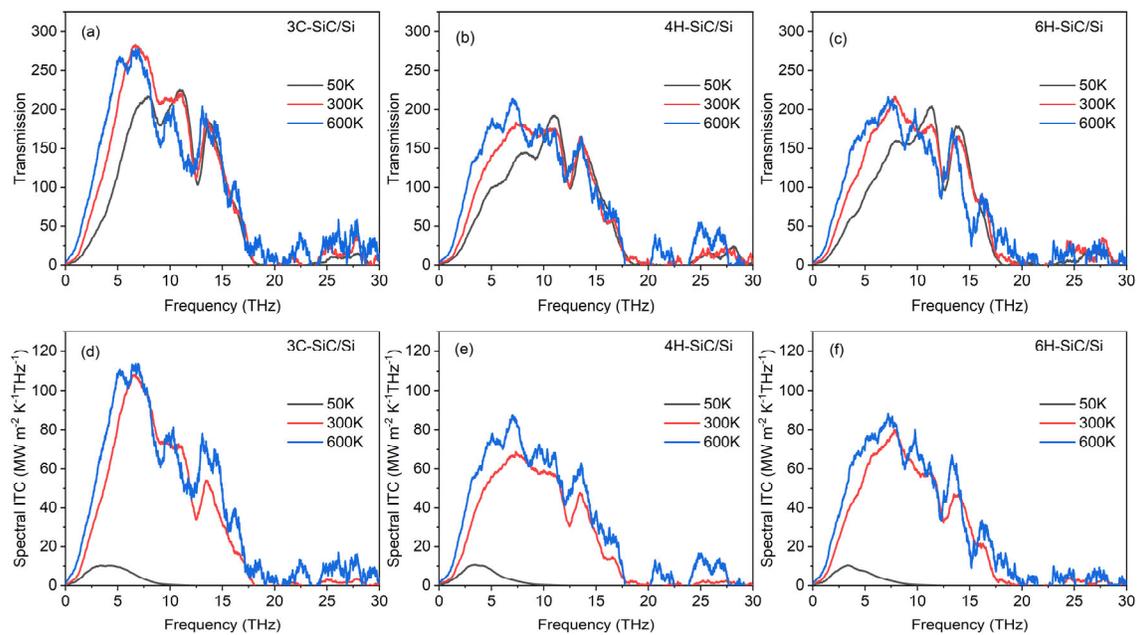

**Figure 4.** The transmission coefficient of (a) 3C-SiC/Si, (b) 4H-SiC/Si and (c) 6H-SiC/Si interfaces and the spectral ITC after quantum correction of (d) 3C-SiC/Si, (e) 4H-SiC/Si, and (f) 6H-SiC/Si interfaces.

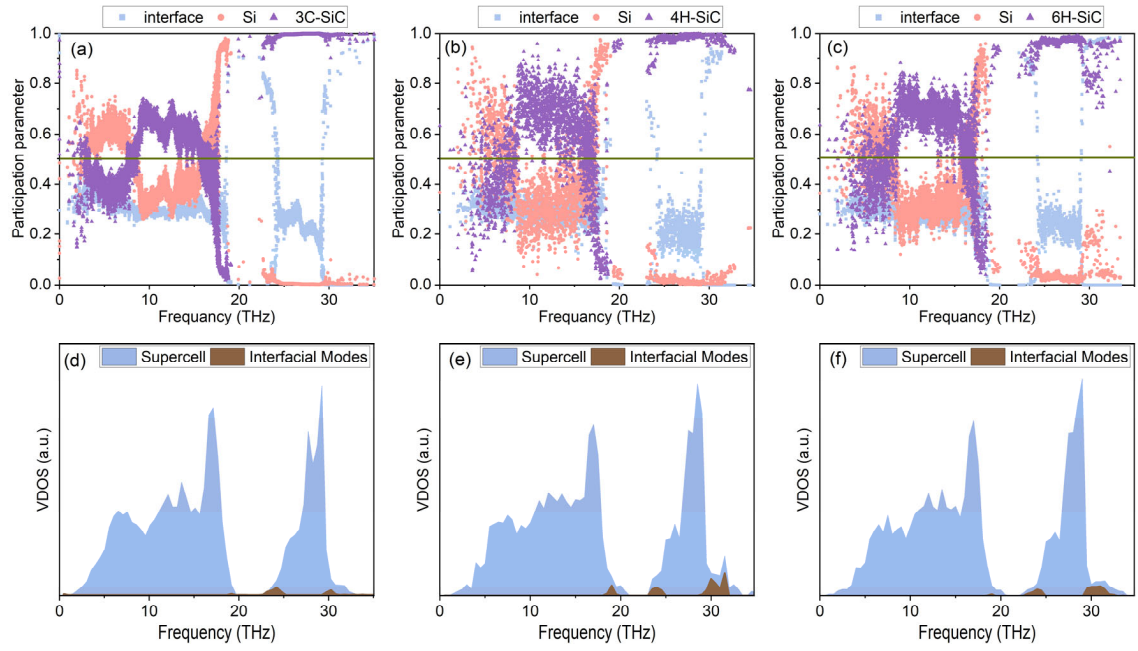

**Figure 5.** The participation parameter contribution of (a) 3C-SiC/Si, (b) 4H-SiC/Si, and (c) 6H-SiC/Si interfaces, and the VDOS of (d) 3C-SiC/Si, (e) 4H-SiC/Si, and (f) 6H-SiC/Si interfaces calculated by lattice dynamics.

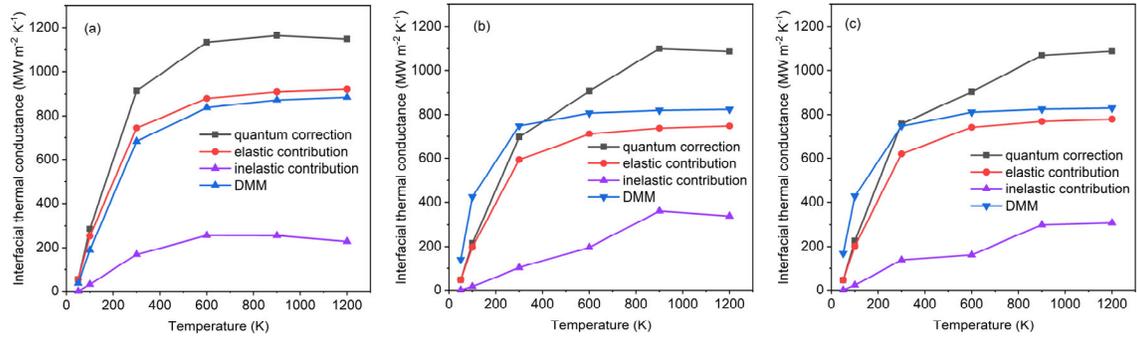

**Figure 6.** The corrected ITC contributed from the elastic and inelastic scatterings of (a) 3C-SiC/Si, (b) 4H-SiC/Si and (c) 6H-SiC/Si interfaces. The ITC contributed by the elastic scatterings is also calculated using the diffuse mismatch model (DMM).